\documentclass[%
 aip,
 amsmath,amssymb,
 reprint,
]{revtex4-1}

\usepackage[english]{babel} 
\usepackage{graphicx}
\usepackage{dcolumn}
\usepackage{bm}

\usepackage[utf8]{inputenc}
\usepackage[T1]{fontenc}
\usepackage{mathptmx}
\usepackage{etoolbox}

\makeatletter
\def\@email#1#2{%
 \endgroup
 \patchcmd{\titleblock@produce}
  {\frontmatter@RRAPformat}
  {\frontmatter@RRAPformat{\produce@RRAP{*#1\href{mailto:#2}{#2}}}\frontmatter@RRAPformat}
  {}{}
}%
\makeatother
\begin{document}

\preprint{AIP/123-QED}

\title[]{Tailored magnetic domain-wall ratchet effects \\through density gradients in triangular-hole arrays}

\author{Julia Herrero-Albillos}
\email{Julia.Herrero@unizar.es}
\affiliation{\mbox{Instituto de Nanociencia y Materiales de Aragón (INMA), CSIC--Universidad de Zaragoza, Zaragoza 50009, Spain}}
\affiliation{\mbox{Departamento de Ciencia y Tecnología de Materiales y Fluidos, Universidad de Zaragoza, Zaragoza 50018, Spain}}

\author{María Ángeles Laguna-Marco}
\affiliation{\mbox{Instituto de Nanociencia y Materiales de Aragón (INMA), CSIC--Universidad de Zaragoza, Zaragoza 50009, Spain}}
\affiliation{\mbox{Departamento de Física de la Materia Condensada, Universidad de Zaragoza, Zaragoza 50009, Spain}}

\author{María Vélez}
\affiliation{\mbox{Departamento de Física, Universidad de Oviedo, 33007 Oviedo, Spain}}
\affiliation{\mbox{Centro de Investigación en Nanomateriales y Nanotecnología (CINN, CSIC--Universidad de Oviedo), 33940 El Entrego, Spain}}

\author{Xabier Zubizarreta-Iriarte}
\affiliation{\mbox{Instituto de Nanociencia y Materiales de Aragón (INMA), CSIC--Universidad de Zaragoza, Zaragoza 50009, Spain}}

\author{Daniel Arechavaleta}
\affiliation{\mbox{Universidad de Zaragoza, Zaragoza 50009, Spain}}

\author{Aurelio Hierro-Rodriguez}
\affiliation{\mbox{Departamento de Física, Universidad de Oviedo, 33007 Oviedo, Spain}}
\affiliation{\mbox{Centro de Investigación en Nanomateriales y Nanotecnología (CINN, CSIC--Universidad de Oviedo), 33940 El Entrego, Spain}}

\author{Carlos Quirós}
\affiliation{\mbox{Departamento de Física, Universidad de Oviedo, 33007 Oviedo, Spain}}
\affiliation{\mbox{Centro de Investigación en Nanomateriales y Nanotecnología (CINN, CSIC--Universidad de Oviedo), 33940 El Entrego, Spain}}

\author{Michael Foerster}
\affiliation{\mbox{ALBA Synchrotron Light Facility, Carrer de la Llum 2--26, 08290 Cerdanyola del Vallès, Barcelona,Spain}}

\author{Cristina Bran}
\affiliation{\mbox{Instituto de Nanociencia y Materiales de Aragón (INMA), CSIC--Universidad de Zaragoza, Zaragoza 50009, Spain}}

\author{Arantxa Fraile-Rodriguez}
\affiliation{\mbox{Departament de Física de la Matèria Condensada, Universitat de Barcelona, Barcelona, Spain}}
\affiliation{\mbox{Institut de Nanociència i Nanotecnologia (IN2UB), Universitat de Barcelona, Barcelona,Spain}}

\author{José Ignacio Martín}
\affiliation{\mbox{Departamento de Física, Universidad de Oviedo, 33007 Oviedo, Spain}}
\affiliation{\mbox{Centro de Investigación en Nanomateriales y Nanotecnología (CINN, CSIC--Universidad de Oviedo), 33940 El Entrego, Spain}}

\date{\today}

\begin{abstract}
We investigate domain wall (DW) dynamics and morphology in amorphous Co$_{0.86}$Zr$_{0.14}$ thin films patterned with engineered, density-graded triangular-hole arrays. Using Kerr microscopy and high-resolution XMCD-PEEM imaging, we demonstrate that DW propagation is governed by the interplay between hole orientation, density gradients, and local magnetostatic interactions. In cooperative configurations, the density gradient reinforces the geometric ratchet effect, whereas in competing configurations, the two mechanisms oppose each other, leading to an inverse ratchet and enabling experimental access to backward DW propagation. Remarkably, density-graded arrays substantially reduce the stochastic variability of depinning fields observed in periodic arrays, resulting in highly reproducible DW motion. The depinning fields are described by a rigid DW model and an empirical scaling law incorporating both inter-hole spacing and array periodicity. These results demonstrate that engineered density gradients provide an effective route to deterministic control of DW propagation, with potential applications in spintronic and neuromorphic devices.

\end{abstract}

\maketitle
\section{Introduction}
Magnetic domain walls (DWs) are fundamental entities in ferromagnetic materials. The ability to precisely control and manipulate their propagation is of central importance for spintronic technologies, including magnetic random-access memories, logic circuits, and neuromorphic computing devices. In thin films, DW motion is strongly influenced by defects and micro-structural inhomogeneities, which act as pinning sites that hinder or redirect domain-wall propagation. In particular, increasing the number of pinning sites enhances the depinning field and, consequently, the coercivity of the film \cite{Kersten1956, Jiles1991, Chikazumi1997}. This understanding has motivated extensive efforts to engineer artificial pinning landscapes that enable controlled and reproducible DW propagation\cite{Adeyeye2008}. Among the various approaches, micro- and nanopatterned arrays of holes have demonstrated particular versatility, as their geometry can be engineered on demand to modulate the pinning strength and control the preferred propagation pathways \cite{CastanGuerrero2013,  Merazzo2012, Heyderman2006, Mallick2018, KRUPINSKI2021}.

\begin{figure}[!t]
 \centering
 \includegraphics[width=0.85\columnwidth]{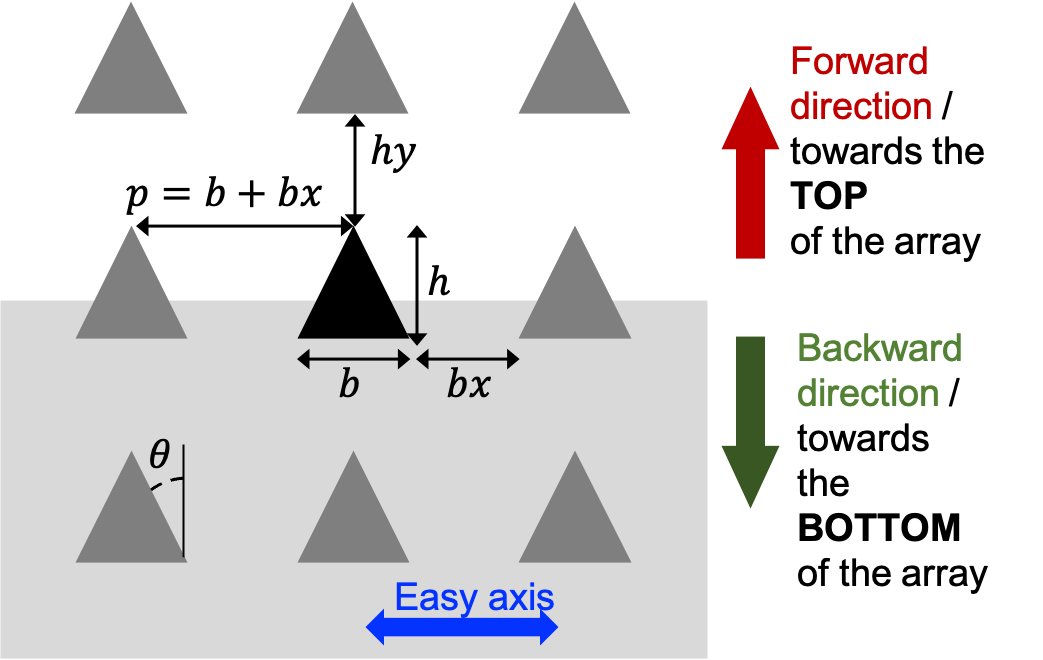}
 \caption{Diagram of part of an array, indicating the relevant parameters: $b$ = base and $h=$ height of the triangles, $bx=$ distance between the bases of the triangles, $hy=$ vertical distance from the tip of one triangle to the base of the upper one. The white and light gray backgrounds represent two different magnetic domains, with the straight line separating them representing a simplified version of the domain wall. The red and green arrows indicate the forward and backward directions respectively, as referred to throughout the paper. The top and bottom of each array are defined according to the triangle orientation: the top edge corresponds to the side where the vertices of the triangles lie at the outermost boundary, while the bottom edge corresponds to the side where the bases lie at the outermost boundary. The blue arrow marks the magnetic easy axes (EA) direction.}
 \label{fig:Param}
\end{figure}

Within these engineered energy landscapes, asymmetric pinning centers are especially relevant because broken spatial symmetry gives rise to magnetic ratchet effects, resulting in preferential DW propagation along a specific direction \cite{PerezJunquera2008, alija_influence_2010, alija_crossed_2011}.
In particular, arrays of triangular holes constitute a simple and robust implementation of this concept. Figure~\ref{fig:Param} schematically illustrates a region of one such array, with a domain wall pinned at a row of triangular holes. This asymmetric geometry results in different depinning fields for forward ($H_F$) and backward ($H_B$) domain-wall propagation, as defined by the triangle orientation \cite{CastanRatchet2014, Herrero2018}, where forward and backward denote domain-wall propagation toward the triangle-tips and triangle-bases, respectively.

Theoretical descriptions of these systems typically treat the DW either as a rigid interface or a deformable elastic line, thereby neglecting its internal magnetic structure \cite{Bertotti1998, Jiles1991}. Within the latter description, the wall energy is largely determined by its length and deformation as it interacts with defects or artificial pinning sites.
In this framework,  the ratio between the forward and backward depinning fields for periodic arrays of triangular holes is predicted to be $H_F/H_B=\sin\theta$, where $\theta$ is the angle between the triangle side and the vertical (see Fig.~\ref{fig:Param})\cite{Marconi2011}. 
Because this ratio is entirely determined by the triangle geometry, the attainable directional asymmetry is intrinsically limited to values corresponding to practical triangle aspect ratios (i.e., with $b \approx h$), since excessively elongated triangles would require impractical use of space within the array.
However, these predictions have not yet been experimentally verified because the geometries investigated to date do not provide experimental access to backward domain-wall propagation.

Moreover, in periodic arrays of this type, the ratchet effect coexists with stochastic depinning processes arising from intrinsic material disorder. As a result, the depinning field exhibits a broad distribution along the array, with dispersions of up to 50$\%$ \cite{Herrero2018}. Such fluctuations do not preclude ratchet operation, provided that the difference between $H_F$ and $H_B$ is sufficiently large. However, because the maximum directional asymmetry is itself limited by the triangle geometry, the ratchet effect becomes highly sensitive to these stochastic fluctuations. Consequently, hole asymmetry alone is insufficient to ensure robust and deterministic domain-wall propagation, limiting both the reproducibility of the response and its practical implementation.

This raises a broader question: Are there additional geometrical parameters that can stabilize —or even reverse— the ratchet effect? A natural parameter to consider is the spatial distribution of the pinning centers. Introducing a density gradient is expected to generate a corresponding gradient in the depinning field, thereby providing an independent source of asymmetry beyond the triangle geometry itself. Such spatial variations in the pinning landscape are also relevant to more complex magnetic architectures, such as self-organized magnetic networks, where local geometrical variations can naturally generate heterogeneous pinning landscapes \cite{Chen2025}. Here, for the first time,  we investigate the effect of hole density gradients combined with shape asymmetry. The relative orientation of these two asymmetries defines two configurations. In the \textit{cooperative configuration}, both favor DW propagation in the ratchet (forward) direction. In the \textit{competing configuration}, the density gradient counteracts the shape-induced asymmetry, favoring DW propagation in the backward direction.

In this work, we present a systematic study of domain wall dynamics in two-dimensional arrays of asymmetric holes, addressing the combined influence of triangle orientation, pinning-site density gradient, and triangle size. We first investigate periodic arrays in amorphous Co$_{0.86}$Zr$_{0.14}$ films, characterized by cross-tie domain walls, before introducing pinning-site density gradients to separate the effects of local hole asymmetry and global geometrical modulation. This enables us to identify how these two mechanisms cooperate or compete to determine domain wall pinning, propagation, and the resulting ratchet effect.
 
\section{Array geometry}
Since the wall--hole pinning interaction is expected to be strongest when the characteristic size of the pinning sites is comparable to the domain wall width, the triangular holes were designed to match the characteristic width of the cross-tie domain wall tails (approximately 2~$\mu$m in amorphous Co$_{0.86}$Zr$_{0.14}$ films \cite{perez-junquera2006}). Accordingly, fifteen two-dimensional arrays of triangular holes were patterned, with the base $b$ ranging from 2.5 to 6~$\mu$m. In all arrays, the triangle height was set to $h=b$, yielding a constant $\sin\theta=1/\sqrt{5}$ across all geometries considered. This constitutes a natural scaling of our previous work on Co$_{0.7}$Si$_{0.3}$ films, where triangular holes of 0.5 and 1~$\mu$m were investigated in the presence of Néel domain walls with a characteristic width of about 400~nm, thereby preserving a comparable ratio between the characteristic domain wall size and the artificial pinning sites \cite{Herrero2018}, while extending the study to asymmetric pinning of cross-tie domain walls.

\begin{figure}[!b]
\centering
\includegraphics[width=8.5cm]{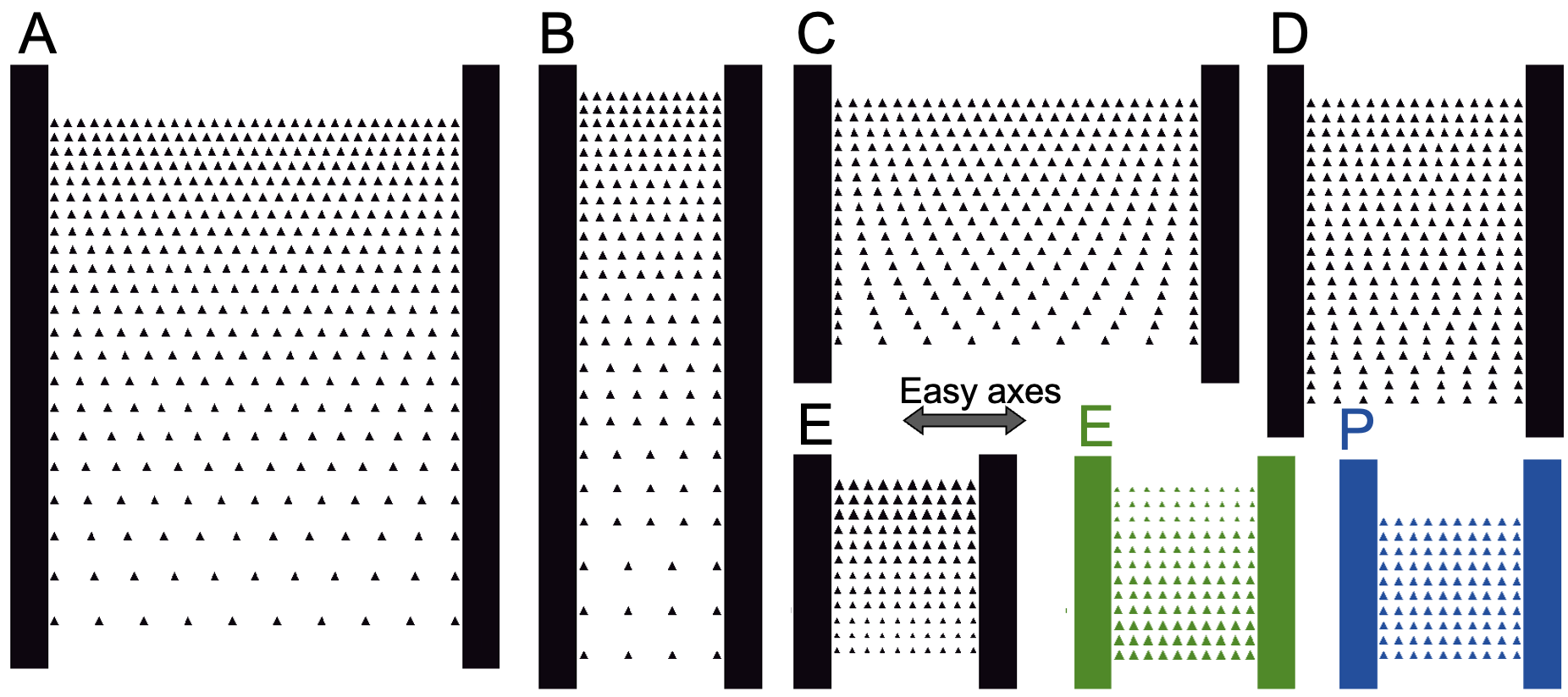}
\caption{Representative array layouts. A- to D-type arrays: all triangles have the same size. Only the configurations in which the triangle orientation and the gradient in triangle density cooperate to enhance the ratchet effect are shown. The corresponding arrays, in which these two contributions compete, were also fabricated but are not displayed. E-type arrays: arrays with a gradient in triangle size. Depicted in black, the triangle orientation and the size gradient cooperate, whereas in green they compete. P-type arrays: periodic array with constant $b$ and $x=y=1$, in which the ratchet effect only arises from the triangle orientation.}
\label{fig:Arrays}
\end{figure}

The spacing of the triangles along their bases is denoted by $bx$, and the separation between rows by $hy$, both varying between 2.5 and 25~$\mu$m. The arrays were delimited by trenches, spaced 1.25 to 2.5 $\mu$m from the holes. The magnetization easy axis was oriented parallel to the base of the triangular holes. The geometrical parameters are defined in Fig.~\ref{fig:Param}, while the complete array geometries are shown in Fig.~\ref{fig:Arrays}, where the different array types, hereafter referred to as A- to E and P-type arrays, are introduced.

Two distinct strategies were followed to introduce a gradient in the available magnetic area: by varying the triangle density while keeping the triangle size constant (A- to D-type arrays), and by varying the triangle size while keeping the triangle density constant (E-type arrays). In the former, two types of local environments were defined around each triangle: in some cases, both the row and column spacings were varied equally ($bx=hy$; A- and B-type arrays); in  others, only the column spacing was varied, while the row spacing was kept constant and equal to $b$ (C- and D-type arrays). In some arrays, the density variation was introduced in every row (A- and C-type arrays), whereas in others, it was applied every three rows (B-, D- and E-type arrays). The decrease in the available magnetic area was oriented either along the direction defined by the triangle orientation (A- to E-type arrays, shown in black in Fig.~\ref{fig:Arrays}) or in the opposite direction (illustrated by the array shown in green). Periodic arrays, i.e., without any gradient, were also fabricated (P-type arrays, shown in blue).

\begin{figure}[tb]
 \centering
\includegraphics[width=0.7\columnwidth]{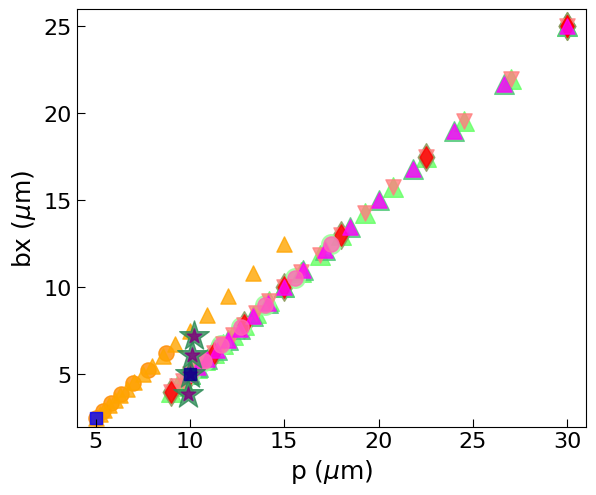}
\caption{Parameter space in terms of $p = bx + b$ and $bx$ explored by the fabricated arrays. Orange markers represent arrays with $b=2.5\,\mu$m, pink/red markers represent arrays with $b=5\,\mu$m, and green markers represent arrays with $b=5\,\mu$m in which the density gradient is opposite to that of the other arrays. Star-shaped markers indicate arrays with variable $b$ while maintaining an approximately constant period, $p \approx 10\,\mu$m. Blue squares represent periodic arrays with constant $b$.}
 \label{fig:FigTabla}
\end{figure}

Table~\ref{tab:variations} summarizes the geometrical parameters of all fabricated arrays. Figure~\ref{fig:FigTabla} shows their distribution within the explored parameter space, highlighting the two main complementary design strategies adopted in this work: arrays with constant triangle size ($b=\mathrm{const.}$) and arrays with approximately constant period ($p\approx\mathrm{const.}$).

\begin{table}[tb]
\caption{Variation of the dimensionless parameter $x$, which determines the inter-triangle distance ($bx$), and of the triangle size $b$ and row spacing ($hy$) for each array type. Arrays with constant $b$ were fabricated with $b=2.5$ or $5~\mu$m.}
\centering
\begin{tabular}{l c c c c}
\hline
\textbf{Type} &\textbf{x} &\textbf{b} & \textbf{hy} & \textbf{Variation}\\ 
\hline
\hline
\textbf{A} &0.8-5 & const. & $bx$ & at each row \\ 
\textbf{B} &0.8-5 & const. & $bx$ & every 3 rows \\
\textbf{C} &1-5& const. & $b$ & at each row \\ 
\textbf{D} &1-2.5& const. & $b$ & every 3 rows \\ 
\textbf{E} &2.4-0.6 & 3-6 $\mu$m & $bx$ & every 3 rows \\ \textbf{P} &1& const. & $b$ & none \\ \hline
\end{tabular}
\label{tab:variations}
\end{table}

\section{Results and discussion}

\subsection{DW Ratchet in Periodic Arrays}

\begin{figure}[!htb]
\centering
\includegraphics[width=0.82\columnwidth]{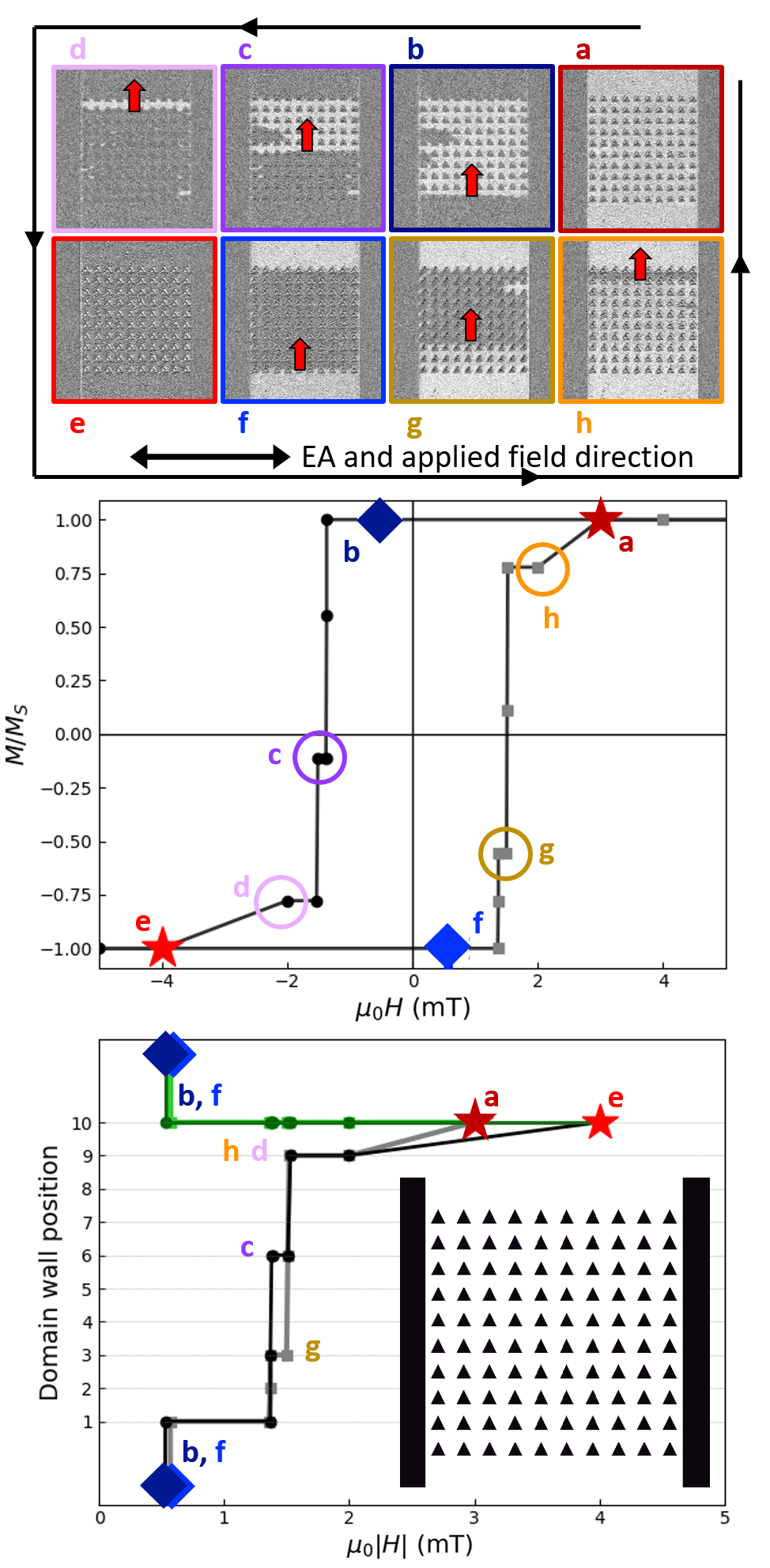}
\caption{Top: Selected Kerr microscopy images of a P-type periodic array with triangle size $b=2.5$ $\mu$m, captured along a magnetic hysteresis loop. The black arrow surrounding the images indicates the sequence along the hysteresis loop. Light-gray and dark-gray contrast in the images corresponds to magnetization parallel and antiparallel to an horizontal positive applied field, respectively. The red arrows indicate the always-upward (forward) motion of the lower domain walls. Kerr images are identified by the same colors and letters as those used in the two graphs below. Center: magnetization as a function of the applied field obtained from the analysis of the images. For clarity, only field values associated with magnetization changes are plotted as full symbols. Red stars  and blue diamonds mark the onset of array saturation and reversal of the continuous film, respectively. Bottom: domain wall position as a function of the absolute value of the applied field for the same hysteresis loop. The forward motion of the lower DW is shown in black (first branch) and gray (second branch), while the static position of the upper DW is shown in dark green (first branch) and light green (second branch). Letters, color coding and symbols are the same as in the hysteresis loop graph. Note that only the bottom domain wall moves upward. The inset shows the array geometry.}
 \label{fig:Periodic}
\end{figure}

Fig.~\ref{fig:Periodic} presents representative Kerr microscopy images recorded at different points of a hysteresis loop together with the corresponding hysteresis loop of a periodic 2.5~$\mu$m-triangular-hole array. 

Starting from the saturation of the entire sample ($a$ in Fig.~\ref{fig:Periodic}), when performing the hysteresis loop, the first event that occurs as the field is reversed is the complete reversal of the magnetization in the continuous film at $\mu_0 H$ $=$ $0.5$~mT ($b$ in Fig.~\ref{fig:Periodic}). At this point, domain walls become pinned at both ends of each array, since the coercive field of the patterned film is higher than that of the continuous region. For this particular array, small reversed domains nucleate near the trenches; however, they do not significantly affect the subsequent overall magnetic behavior of the array.

As the magnetic field is further increased, the domain wall initially pinned at the bottom edge begins to propagate toward the opposite end ($c$ and $d$ in Fig.~\ref{fig:Periodic}); i.e., in the forward direction as defined in Fig.~\ref{fig:Param}. Changes in magnetization within the arrays are then driven by the forward motion of this wall. On the contrary, the other wall—pinned at the top edge—remains essentially stationary. As the applied field increases, the magnetization of the rows progressively reverses until all of them become aligned with the continuous film and the external field, leading to the magnetic saturation of the entire sample ($e$ in Fig.~\ref{fig:Periodic}). Upon field reversal, the process is repeated, but with the magnetization pointing in the opposite direction: two new domain walls form again at both ends of the array ($f$ in Fig.~\ref{fig:Periodic}), and it is the one closer to the bottom end that propagates across the structure, reversing the magnetization while the wall at the top end remains pinned ($g$ and $h$ in Fig.~\ref{fig:Periodic}). 

In the 5~$\mu$m triangular-hole arrays, similar behavior is observed, except that no small reversed domains nucleate near the trenches, and the pinning fields are nearly half those of the smaller triangles.

In these periodic arrays with geometrically identical rows, the ratchet effect originates from the difference between the pinning fields for forward ($H_F$) and backward ($H_B$) domain-wall propagation, which are the same for all rows. 
However, the measured depinning fields span a wide range across the arrays, resulting in step-like features in the hysteresis loops, as shown in the middle panel of Fig.\ref{fig:Periodic}. These features arise from the combination of ratchet and  stochastic effects in domain-wall motion: a wall can only de-pin from a given row once it has reached it from a lower row, and the subsequent depinning event is strongly influenced by Barkhausen noise associated with random defects and inhomogeneities.
As a result, significant differences in the experimentally measured depinning fields can be observed between different rows, with strong cycle-to-cycle variability, despite the periodic geometry. In contrast, micromagnetic simulations of ideal periodic arrays predict essentially the same depinning field for every row, such that once the domain wall de-pins from the first row, it propagates through the array with little or no additional pinning \cite{alija2009}.
The observed behavior is consistent with our previous reports on the ratchet effect in a different material and dimensions \cite{Herrero2018, PerezJunquera2008}, confirming that triangular holes robustly define the preferred direction of domain wall propagation. However, they do not provide deterministic control over the global magnetization reversal of the array, since the depinning field remains dominated by intrinsic disorder. 
\subsection{Magnetostatic interactions and domain wall structure}
\label{sec:XMCD}
\begin{figure}[!b]
\centering
\includegraphics[width=6.5cm]{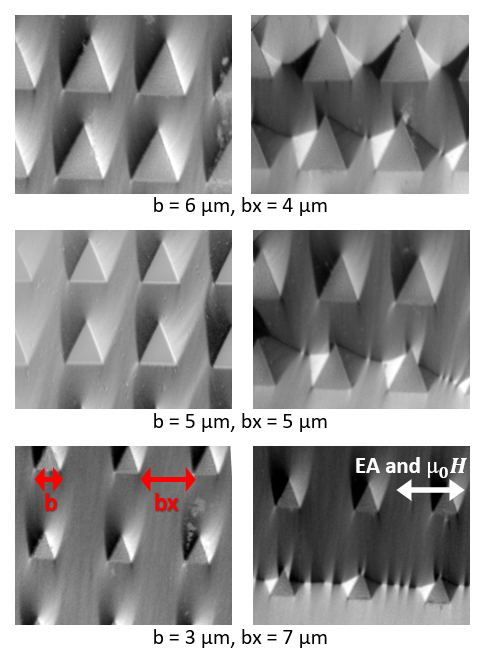}
\caption{XMCD-PEEM images showing the magnetic configuration around triangular holes without (left) and with (right) a domain wall, for different geometrical parameters: $b = 6$ $\mu$m, $bx = 4$ $\mu$m (top); $b = bx = 5$ $\mu$m (middle); and $b = 3$ $\mu$m, $bx = 7$ $\mu$m (bottom). The magnetization projection corresponds to an angle of $70^\circ$ with respect to the horizontal.}
\label{fig:PEEM}
\end{figure}

In order to understand the details of the DW depinning process, it is fundamental to gain insight into the details of the DW configuration, which has been investigated by XMCD-PEEM. Fig.~\ref{fig:PEEM} shows representative images acquired for different geometrical parameters,  with and without a domain wall pinned between triangles.

Even in the absence of a domain wall, the magnetization around the triangles is highly non-uniform. Near the triangle tips, substantial magnetostatic charge accumulates at the hole edges, whereas near the bases, the magnetization remains largely parallel to the edges, resulting in much smaller charge. This asymmetric magnetostatic landscape results from the interplay between the triangular geometry and the in-plane uniaxial anisotropy, whose easy axis is parallel to the triangle bases. If a domain wall were to approach a triangle row, it would experience this non-uniform magnetostatic charge distribution, in addition to the external field and its own magnetostatic charge, providing a mechanism for the local ratchet effect.

Fig.~\ref{fig:PEEM} also shows that when a domain wall is present between triangles (right panels), its configuration departs significantly from that expected for a simple rigid or elastic wall and its structure has a strong dependence on geometrical parameters. DWs in the unpatterned film are cross-tie type with a density of around 1 vortex-antivortex (V-AV) pairs per $\mu$m (see supplementary material). For the largest hole separations, domain walls pinned between triangles largely retain the continuous film cross-tie configuration (e.g., bottom right panel of Fig.~\ref{fig:PEEM}, $bx = 7~\mu$m). However, the number of V–AV pairs is significantly reduced with respect to the unpatterned film, with only 2–3 pairs observed between triangles instead of the expected $\thicksim7$. For smaller separations between triangles (e.g., middle and top right panels of Fig.~\ref{fig:PEEM}, $bx = 5$ and $4~\mu$m), this effect becomes more pronounced: the cross-tie structure is further suppressed, and in some cases, only one V–AV pair (or one vortex and 2 half-antivortices) is observed. Interestingly, a similar reduction in the number of V/AV pairs has been reported for cross-ties in the presence of transverse fields \cite{crosstie_Htrans}, suggesting that the observed behavior may be associated with the transverse magnetostatic fields generated by the patterned triangles.

\begin{figure}[!b]
\centering
\includegraphics[width=7cm]{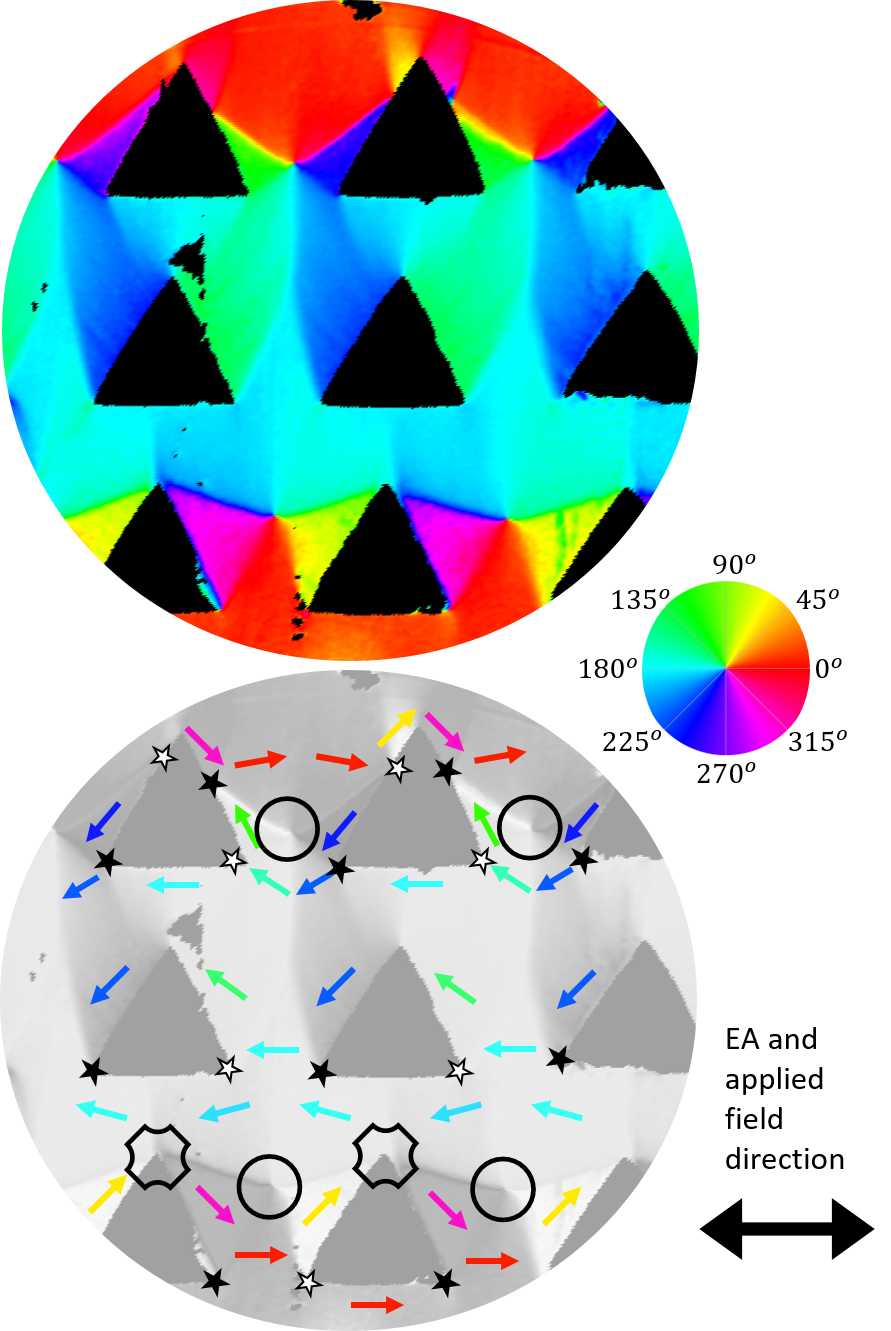}
\caption{Top: Direction of the in-plane magnetization  at the top of an array with $b = 6$ $\mu$m, with a domain wall pinned at the top edge of the array and another one two rows below. Black areas correspond either to the patterned triangles, where no magnetic material is present, or to superficial defects that locally suppress the PEEM signal. Bottom: Same in-plane magnetization  map in gray scale with the position of vortices (circles), antivortices (concave squares) and half-antivortices (stars). The colored arrows represent the averaged direction of the magnetization in their surrounding area.}
\label{fig:2Dmap}
\end{figure}

The map of in-plane magnetization depicted in Fig.\ref{fig:2Dmap} shows two domain walls in an array, one pinned at the top edge and another one two rows below. The map enables a more detailed analysis of the pinned DWs, which can be regarded as composite domain walls combining the V/AV pairs within the cross-tie with spin textures at the edges of the patterned triangles. A general property of a hole within an in-plane magnetized thin film is the presence of a pair of edge spin textures with fractional winding number $q=-1/2$ (i.e. a pair of half-antivortices) \cite{Tchernyshyov2005, Rodriguez-Rodriguez2008}, corresponding to singular points at two opposite sides of the hole where the magnetization curls around the hole in opposite senses.
Throughout the array shown in Fig.\ref{fig:2Dmap}, these edge spin textures are located at the two corners of the base of each triangular hole.
The white and black stars indicate the one-in/two-out and two-in/one-out magnetic configurations, respectively, with their positions interchanging depending on the overall positive or negative horizontal magnetization of the surrounding domain. 

For the DW pinned at the uppermost row of the array, we find an additional pair of half antivortices that accommodate the magnetization reversal within the closure domains surrounding the triangular hole. In the inter-hole region, a single vortex texture appears at the coalescence point of neighboring closure domains (see circles in Fig.\ref{fig:2Dmap}). Thus, this pinned DW has adopted a three spin texture configuration (two $q=-1/2$ half-antivortices surrounding a central $q=+1$ vortex), similar to vortex walls found in patterned strips \cite{Clarke2008}.  

The situation is different for the lower DW. In this case, the DW has progressed along the triangular holes just until the triangle tips; and displays a clear cross-tie configuration with V/AV distance matching the triangular array periodicity: there is a vortex in the inter-hole region and an antivortex (see concave squares in Fig.~\ref{fig:2Dmap}) at each triangle tip (with winding number $q=-1$ due to the two-in/two out configuration of the closure domains at the triangle tip). 
An analysis of the spin textures of the cross-ties in the lower DWs shown in Fig.~\ref{fig:PEEM} reveals the same behavior as that observed in Fig.~\ref{fig:2Dmap}:  the cross-tie DWs host $n$ V textures and contain only $n-1$ AVs while the remaining $q=-1$ winding number contribution is provided by a pair of $q=-1/2$ half-antivortices. As DWs propagate in the forward direction, the half antivortices slide along the triangle edges until they recombine into a single antivortex at the triangle tip and, consequently, restore the cross-tie configuration needed for propagation through the free space in between the rows of triangular holes.

This evolution of the composite domain wall upon crossing the pinning sites is also typical in patterned strips \cite{Clarke2008,Wuth2013} and can have a direct impact on the switching field distribution through the variation in domain wall energy. Moreover, in this work, the array of holes imposes additional constraints due to the need to match V/AV distance with the relevant array dimensions, namely $bx$ at the base and $p=b+bx$ at the tips of the triangles. This is also expected to impact domain-wall propagation when the separation between triangles varies from row to row in non-periodic arrays with graded periodicity.

\subsection{Beyond periodic arrays: Tailoring domain wall pinning and ratchet asymmetry}
In non-periodic arrays of triangular holes, domain wall propagation is governed by two concurrent mechanisms: the intrinsic asymmetry imposed by the triangular shape and the variation in hole density along the vertical direction. The triangular orientation defines the forward direction and locally favors forward propagation through the asymmetric magnetostatic landscape seen in Figs.~\ref{fig:PEEM}~and~\ref{fig:2Dmap}. The hole density gradient modulates the available magnetic length (i.e. distance between holes) and thus the local coercive field, with higher hole densities leading to shorter effective lengths and larger $H_c$ \cite{Chikazumi1997, alija2009, Askey2024, pirota_coercive_2008, RUIZFEAL2002, prieto_fabrication_2008, Perez-Junquera2004, Marconi2011}.   

Therefore, depending on whether these two mechanisms act in the same or in opposite directions within an array, two distinct cases can be distinguished.  We refer to arrays as \textit{cooperative} when the orientation of the triangular holes and the density gradient both favor forward propagation, thereby enhancing the ratchet effect (black A to E-type arrays in Fig.~\ref{fig:Arrays}).  Conversely, in \textit{competing} arrays, the triangles orientation and the density gradient oppose each other (green E-type array in Fig.~\ref{fig:Arrays} and similar configurations).
The specific magnetic behavior of each case is analyzed below.

\subsubsection{Phenomenological description of cooperative (ratchet-enhancing) arrays}

\begin{figure}[!b]
\centering
\includegraphics[width=0.95\columnwidth]{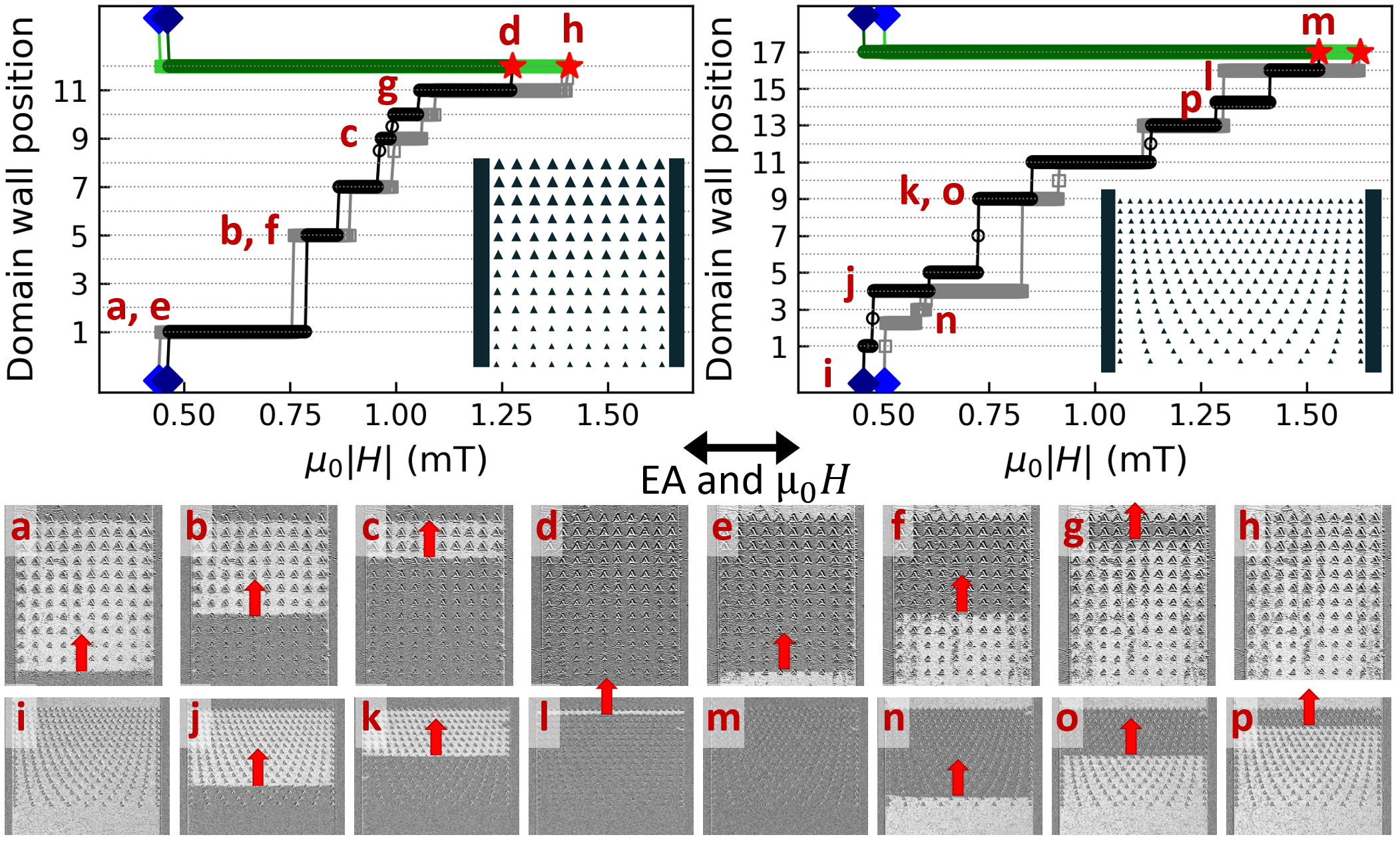}
\caption{Top: Field dependence of the domain-wall position for a cooperative E-type array with variable $b$ (left) and a cooperative C-type array with $b = 2.5~\mu$m (right). The forward motion of the lower domain wall is shown in black (first branch) and gray (second branch), while the static position of the upper domain wall is shown in dark green (first branch) and light green (second branch). Red and blue stars mark the onset of array saturation and reversal of the continuous film, respectively. The insets show the array geometries. Bottom: Kerr images corresponding to selected points along both hysteresis loops, identified by the letters indicated in the hysteresis loops and the images. The red arrows indicate the always-upward (forward) motion of the lower domain walls.}
 \label{fig:DWpossCol}
\end{figure}

The global magnetic behavior of the magnetization  during a hysteresis loop closely resembles that of the periodic case: DWs become pinned at both ends of the array at $\mu_0 H = 0.5 \pm 0.1$~mT, and subsequent changes in the magnetization are governed by the forward only propagation of DWs within the array. The unidirectional behavior, shown in Fig.~\ref{fig:DWpossCol} for two arrays, is observed across all cooperative arrays regardless of their specific geometric parameters.

It is worth noting that the forward depinning fields of the first and penultimate rows display a substantially larger dispersion over repeated hysteresis cycles than those of the central rows. Propagation from the first row into the array requires the annihilation of vortex–antivortex pairs, whereas at the penultimate row the propagating wall interacts repulsively with the opposing domain wall located at the top of the array. For this reason, depinning events associated with these two rows are excluded from the analysis below.

\subsubsection{Phenomenological description of competing (ratchet-suppressing) arrays }

\begin{figure}[!t]
\centering
\includegraphics[width=0.95\columnwidth]{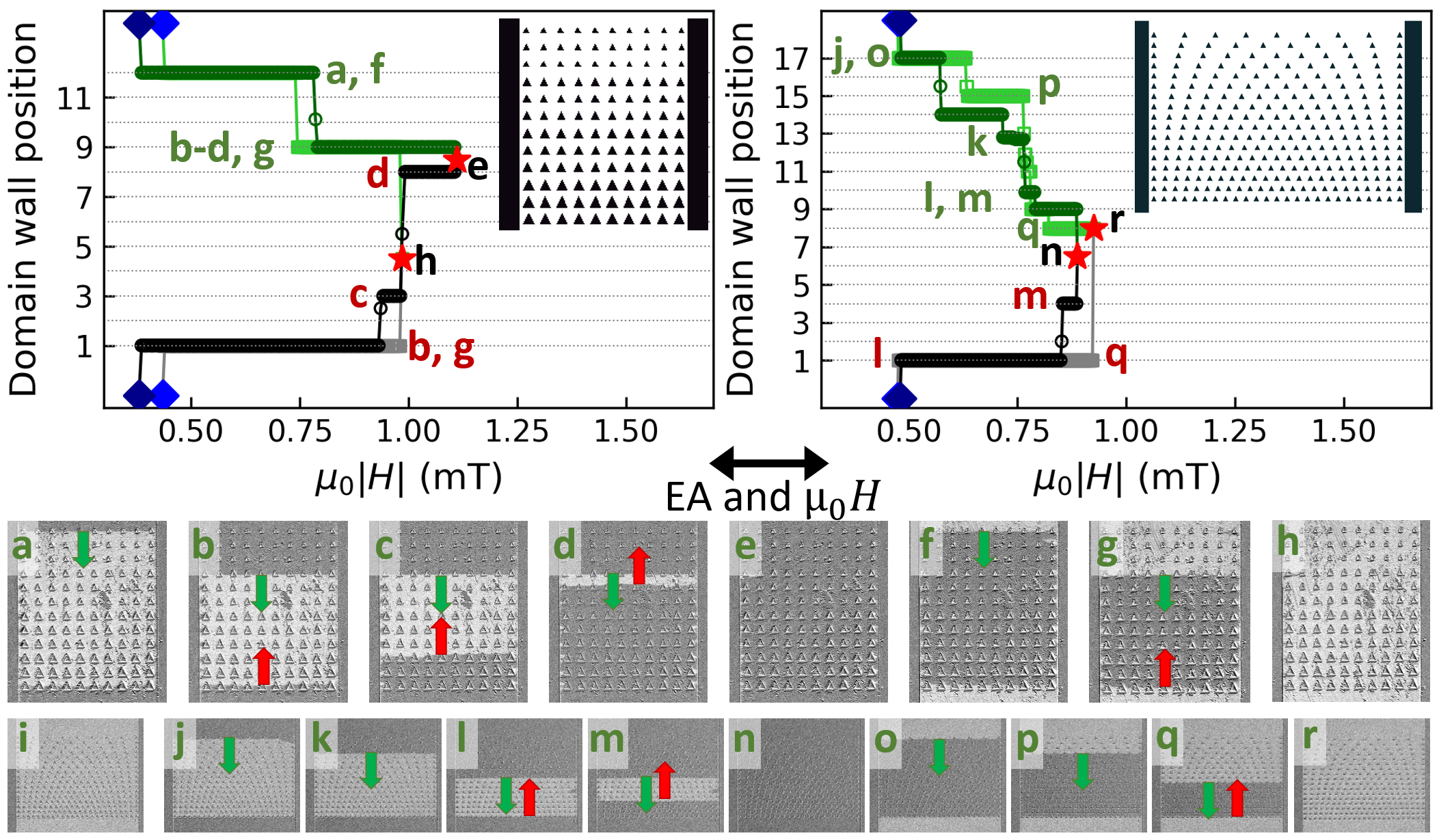}
\caption{Top: Field dependence of the domain-wall position for a competing E-type array with variable $b$ (left) and a competing C-type array with $b = 5~\mu$m (right). The forward motion of the lower domain wall is shown in black (first branch) and gray (second branch), while the backward motion of the upper domain wall is shown in dark green (first branch) and light green (second branch). Red and blue stars mark the onset of array saturation and reversal of the continuous film, respectively. The insets show the array geometries. Bottom: Kerr images corresponding to selected points along both hysteresis loops, identified by the letters indicated in the hysteresis loops and the images. The red arrows indicate the upward (forward) motion of the lower domain walls, and the green arrows indicate the downward (backward) motion of the upper domain walls.}
\label{fig:DWpossComp}
\end{figure}

In the case of hysteresis loops for the competing arrays, as shown in Fig.~\ref{fig:DWpossComp}, domain walls are also initially pinned at both ends of the array. However, in these arrays, it is the domain wall pinned at the top that propagates first, in the backward direction. Subsequently, depending on the specific geometry of the array, propagation in the forward direction of the second wall (initially pinned at the bottom end of the array) may occur. At some point—typically when the upper wall is located near the center of the array—a sudden and complete reversal of the array’s magnetization is observed, without the possibility of identifying which of the two walls is responsible for this final switching event. This reversal can occur either after the bottom wall has already propagated in the forward direction, or while that wall still remains pinned at the bottom of the array.

The origin of the domain wall motion in the backward direction—i.e., opposite to the orientation of the triangles—relies on the fact that the density of holes (or, alternatively,  the hole size) is significantly lower at the top end of the array. As a result, based solely on density considerations, the local coercive field is much lower at the top than at the bottom of the array. This preferential propagation remains valid even when accounting for the asymmetry in the driving force for domain wall motion imposed by the orientation of the triangular holes.

It is worth noting that this behavior differs from the previously reported \textit{crossed ratchet effect} \cite{PerezJunquera2008, alija2009}, where backward domain wall motion originates from the asymmetric depinning of kinks and anti-kinks propagating perpendicularly to the wall. In our case, the backward propagation results from a coercivity gradient imposed by the density variation along the array, and takes place despite the intrinsic asymmetry of the triangular holes, which by itself would favor forward motion. For a range of parameters of our arrays, the effect of the density gradient dominates over the shape-induced asymmetry at the edges of the arrays. This scenario is particularly fortunate, as it provides experimental access to backward propagation mechanism in arrays with asymmetric holes, not experimentally available in prior studies.

\subsubsection{Comparative phenomenology of cooperative and competing arrays}

\begin{figure}[!t]
\centering
\includegraphics[width=6.5cm]{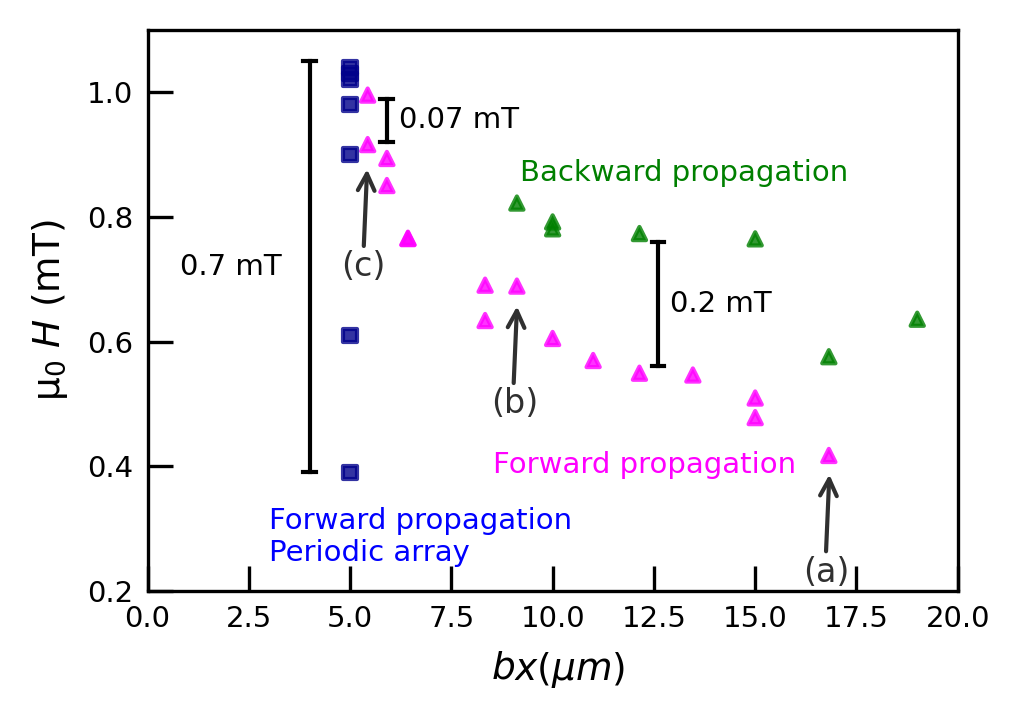}
\caption{Depinning field as a function of triangle separation for three arrays with $b=5\,\mu$m. The pink symbols show the forward depinning field of a cooperating C-type array, the green symbols the backward depinning field of a competing C-type array, and the blue symbols the depinning field of a periodic array (P-type array). (a) Lower forward depinning event of the cooperating array. (b) Higher backward depinning event of the competing array. (c) Higher forward depinning event of the cooperating array. The values $bx$ correspond to the nominal lithographic design, and the field resolution is determined by the acquisition step, ranging from 0.002 to 0.005 mT, resulting in error bars smaller than the data markers.
}
\label{fig:CamposCyG}
\end{figure}

Fig.~\ref{fig:CamposCyG} shows the dependence of the de-pinning fields as a function of the row-dependent triangle separation ($bx$) for two C-type arrays with constant row spacing and triangle size $b=5~\mu$m, together with a periodic P-type array with the same triangle size. The two C-type arrays have mirrored density gradients along the vertical direction so that one is a cooperative array while the other a competing array. For each value of the triangle separation, the data include the forward depinning field from the periodic and cooperative arrays and the backward depinning field from the competing array, enabling a direct comparison for identical geometrical parameters.

In the C-type arrays, the de-pinning fields decreases with increasing triangle separation, as expected. Despite this common trend, the backward and forward depinning fields remain separated by approximately 0.1 to 0.2 mT over the entire range investigated (i.e. being $H_B$  $15\%-50\%$ higher than $H_F$).  Both the forward and backward depinning fields in these arrays exhibit a finite dispersion (about 0.07 mT, see fig.~\ref{fig:CamposCyG}) arising from the Barkhausen noise discussed above. However, its magnitude is approximately one order of magnitude smaller than that observed in the periodic array shown in the figure and remains two to three times smaller than the difference between the backward and forward depinning fields.

The reduced stochasticity observed in these density-graded arrays originates from the systematic variation of hole density, which fixes the available magnetic length between holes and thereby makes the depinning process far more reproducible than in periodic arrays \cite{Herrero2018,PerezJunquera2008,Mallick2018,KRUPINSKI2021}, highlighting the ability of engineered pinning gradients to achieve deterministic control of domain-wall motion.

\subsection{Analysis - Beyond the Kersten model: pinning in asymmetric hole arrays} 

The low dispersion of the measured depinning fields, despite Barkhausen noise, enables systematic comparison across all the investigated arrays and allows the depinning fields to be related to the local parameters $b$, $x$, and $y$ within the arrays.
The resulting dependence can be analyzed in light of two simplified approaches to domain-wall pinning: elastic and rigid domain-wall descriptions. In what follows, these are adapted to the geometry of the triangular hole arrays studied here and compared with the experimental depinning fields measured across the different arrays.

\subsubsection{Elastic and rigid domain-wall models}

The energy change associated with the displacement of a domain wall by a distance $\Delta z$ along the propagation direction is $\Delta E_{DW}=2M_s\mu_0HL_{DW}t \Delta z$, where $L_{DW}$ and $t$ are the domain wall length and film thickness, respectively \cite{Jiles1991}. The corresponding driving field can therefore be expressed as
\begin{equation}
\mu_0H=\frac{1}{2M_sL_{DW}t}\frac{\partial E_{DW}}{\partial z}\label{eq:Zeeman}
\end{equation}

On the other hand, since the seminal work of M. Kersten \cite{Kersten1956}, it has been established, both theoretically and experimentally, that the critical depinning field—and thus the coercivity—in systems with impurities or artificially induced inclusions (i.e., holes) scales with  pinning site density \cite{Chikazumi1997}.  When the inclusions have finite size, the relevant parameter has been shown to be not the center-to-center spacing but the available magnetic length between neighboring pinning sites \cite{pirota_coercive_2008, RUIZFEAL2002, prieto_fabrication_2008, CastanGuerrero2014, Perez-Junquera2004}.

Considering the specific case of asymmetric pinning sites, the role of array geometry on backward/forward depinning fields can be understood within two different simplified frameworks depending on DW surface energy $\sigma$. 

For small $\sigma$, the DW can be considered as an elastic string pinned at the triangle bases to minimize its total length, $L_{DW}=bx$. Following the model developed by Marconi et al. \cite{Marconi2011}, the DW bends under the applied field until depinning occurs, yielding the following estimated depinning fields for our array geometries:
\begin{gather}
    \mu_0H_F^{elastic} = {\frac{1}{2M_S}} {\frac{\sigma}{bx}}\sin\theta \label{eq:HFE}
    \\
    \mu_0H_B^{elastic} = {\frac{1}{2M_S}} \frac{\sigma}{bx}\label{eq:HBE}
\end{gather}

For large $\sigma$, the DW can be considered a rigid line with energy $E_{DW}=\sigma L_{DW} t$, where both the DW length and its surface energy may vary as the wall propagates across the triangles. Coercivity occurs when the applied field matches the local DW energy gradient \cite{Chikazumi1978,alija2009} which for a DW pinned at the triangle basis corresponds to
\begin{equation}
    \frac{\partial E_{DW}}{\partial z}=\sigma t \frac{\partial L_{DW}}{\partial z}+L_{DW} t \frac{\partial \sigma}{\partial z}=\sigma t \frac{\partial L_{DW}}{\partial z}+ b x t \sigma'
    \label{eq:DWEgradient}
\end{equation}
where $\sigma'$ accounts for local variations in the DW surface energy. At the triangle base, the local DW surface energy is denoted by $\sigma_0$. Its variation around this point depends on the direction of propagation, so $\sigma'$ may take different values for forward and backward propagation, denoted by $\sigma'_{0,F}$ and $\sigma'_{0,B}$, respectively.

\begin{figure}[!b]
    \centering
    \includegraphics[width=0.6\linewidth]{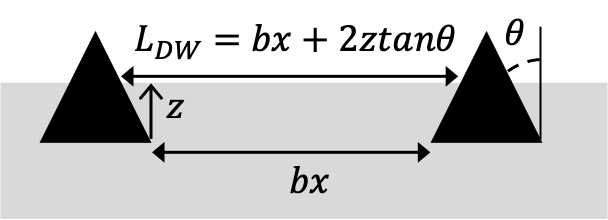}
    \caption{A DW pinned at an intermediate position $z$ along the height of the triangles, with a domain-wall length $L_{DW}= bx+2z\tan \theta$}
    \label{fig:DWrigid}
\end{figure}

For forward propagation, as the wall moves towards the tips, the DW length, as shown in Fig.~\ref{fig:DWrigid}, increases according to $\Delta L_{DW}=bx+2\Delta z\tan\theta$.
Therefore, combining eqs.~\ref{eq:Zeeman} and~\ref{eq:DWEgradient}, the de-pinning fields for forward propagation of a rigid DW can be expressed as

\begin{equation}
    \mu_0H_F^{rigid} = 
    \frac{1}{2M_S} \left(\frac{2\sigma_0\tan\theta }{bx}+\sigma_{0,F}'\right)\label{eq:HFR}
\end{equation}

For backward depinning from the triangle basis, the sharp increase in domain wall length ($\Delta L_{DW}=b$) in a very short spatial interval ($\Delta z\approx \delta_{DW}$ with $\delta_{DW}$ an effective DW width) results in a larger backwards coercivity \cite{alija_crossed_2011}
\begin{equation}
    \mu_0H_B^{rigid} = \frac{1}{2M_S} \left( \frac{\sigma_0 }{bx} {\frac{b}{\delta_{DW}}} + \sigma_{0,B}'\right) \label{eq:HBR}
\end{equation}
as long as $\delta_{DW}<b/2\tan\theta$. 

Whether the domain wall is treated as rigid or elastic, both models highlight the relevance of the inter-hole spacing $bx$ for the local depinning field and of the triangle shape, characterized by the angle $\theta$, for the forward/backward asymmetry.

\subsubsection{Experimental depinning fields and comparison with the models}

The arrays designed for this work enable the experimental investigation of the combined effects of density gradients and directional asymmetry, as well as backward domain-wall propagation, building on these simplified descriptions of local pinning.

\begin{figure}[hb!]
\centering
\includegraphics[width=7cm]{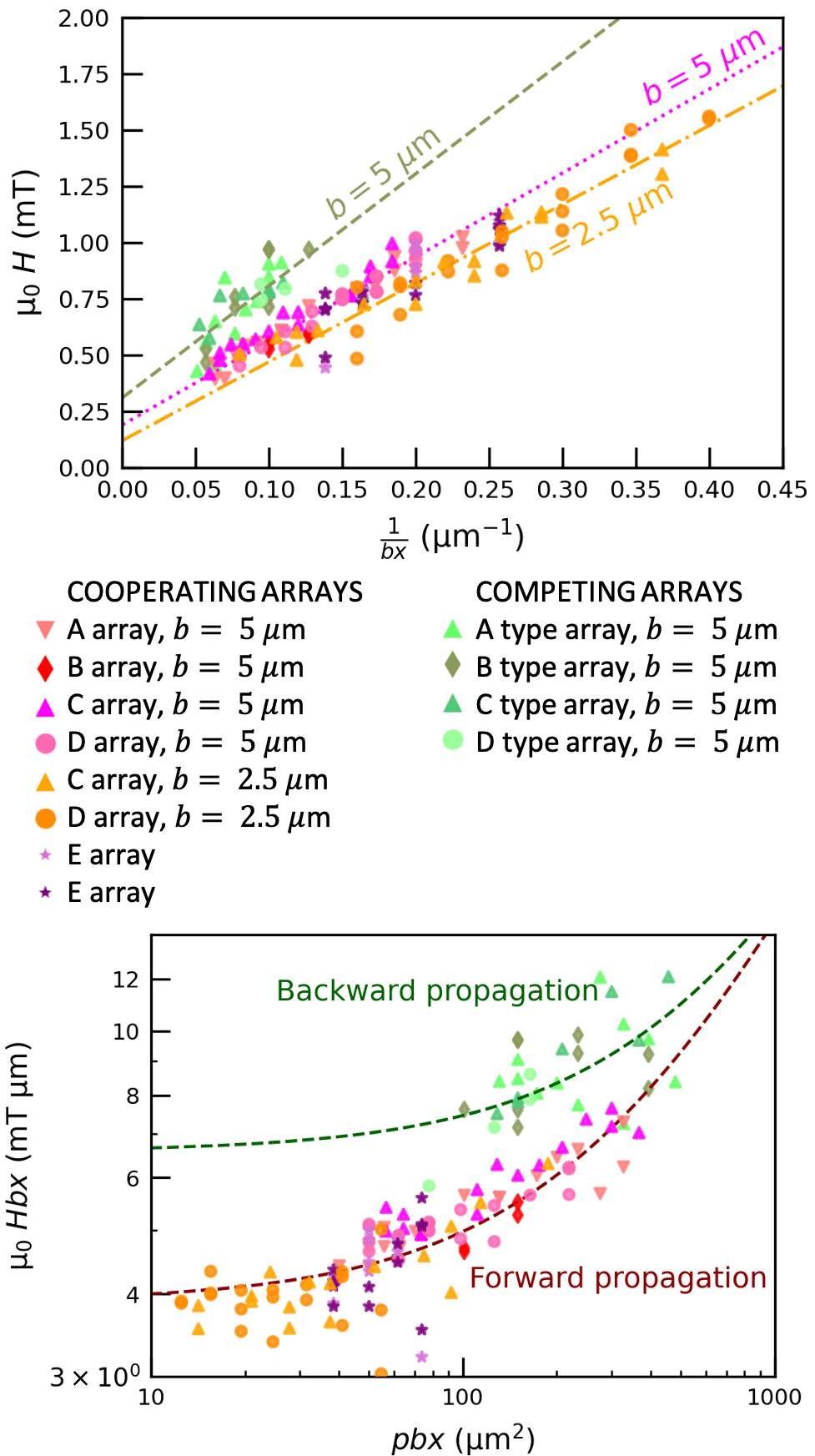}
\caption{Depinning field versus the reciprocal of the horizontal triangle separation for the arrays indicated in the figure legend, corresponding to one hysteresis loop for each array. Warm colors are used for cooperative arrays, green shades for competing arrays, and blue for periodic arrays. Linear fits were performed separately for cooperative arrays ($b=2.5$ $\mu$m in orange and $b=5$ $\mu$m in pink) and competing arrays ($b=5$ $\mu$m in green); data from arrays with variable 
$b$ (marked with stars) and from periodic arrays were excluded from the fits. 
Bottom: The same experimental data are shown, now plotting the product of the depinning field and horizontal triangle separation versus the product of horizontal triangle separation and array periodicity, with both axes displayed on a logarithmic scale. Dashed lines correspond to linear fits in the log–log scale, performed separately for cooperative and competing arrays. Each fit includes all arrays of the corresponding type, regardless of the value(s) of $b$. Labels for the color and symbol coding correspond to those defined in Table~\ref{tab:variations} and Fig.~\ref{fig:Arrays}. The values of $p$ and $bx$ correspond to the nominal lithographic design, and the field resolution is determined by the acquisition step, ranging from 0.002 to 0.005 mT, resulting in error bars smaller than the data markers.
}

\label{fig:Pinningfield}
\end{figure}

Analysis of both cooperative and competing arrays reveals that, in agreement with eqs.~\ref{eq:HFE}, \ref{eq:HBE},~\ref{eq:HFR}~and~\ref{eq:HBR}, the de-pinning fields is primarily governed by the horizontal spacing between adjacent triangles at their bases ($bx$), which determines the effective length of the pinned domain wall when entering a row from the bottom. In particular, as shown in the top panel of Fig.~\ref{fig:Pinningfield}, the depinning field of the constant-$b$ arrays exhibits a linear dependence on $1/bx$, with three distinct trends of the form $H_c = C/bx + D$: two for forward-propagating walls in cooperative arrays, corresponding to $b=2.5$ and $5~\mu$m, and one for backward-propagating walls in the competing arrays with $b=5~\mu$m. 

The two fits for the cooperative arrays yield similar values of the slope ($C_F$), independent of triangle size. This behavior is consistent with both elastic and rigid DW models (eqs.~\ref{eq:HFE} and~\ref{eq:HFR}), which, for a fixed triangle angle $\theta$, do not predict a dependence of $C_F$ on $b$. In contrast, the slope obtained for backward propagation ($C_B$) in the competing arrays is only about $30\%$ larger than for forward propagation ($C_B/C_F=1.3$). This increase is substantially smaller than the factor $1/\sin{\theta}=2.2$ predicted by the elastic DW model (eqs.~\ref{eq:HBE} and~\ref{eq:HFE}).
For the rigid DW model, comparison with the experimental slope ratio yields
$\frac{5~\mu\mathrm{m}}{2\tan{\theta}\delta_{DW}}=1.3$, corresponding to an effective DW width of $\delta_{DW}\approx3.8~\mu$m. This value is consistent with the DW widths (including the tails) determined by PEEM (see Supplementary material for details), supporting the description of the DW as entities where their surface energy can vary depending on the wall position along the triangular holes and emphasizing the importance of the domain wall tails in the pinning process \cite{perez-junquera2006}.

In addition, these different $H_c~ vs.~ 1/bx$ trends include clear non-zero intercepts $D$ with different magnitudes depending on the triangle size ($D$ is roughly $60\%$ larger for  $b=5$ $\mu$m than for $b=2.5$ $\mu$m), so that they cannot be simply attributed to the continuous film coercivity in the limit of very large $bx$. 

The elastic DW model does not account for such a finite intercept. Within the rigid DW model, however, a finite intercept can be directly related to forward/backward variation of the DW surface energy (i.e. $D\propto \sigma_{0,F/B}'$ in eqs.~ \ref{eq:HFR}~and~\ref{eq:HBR}). The confinement effects observed in the PEEM characterization of the pinned cross-tie DWs in Section~\ref{sec:XMCD} provide a possible mechanism for such variations.
When the DW progresses from the triangle base to the tip, the V-AV period needs to adapt, at each vertical DW position, to the available distance between the closure domains of the neighboring triangles. Finally, when the DW reaches the top of the triangle, AVs become pinned at triangle tips so that the V-AV period must match the horizontal array periodicity $p=b+bx$. V-AV period has a direct effect in the cross-tie surface energy \cite{Donahue2012}, so that stretching the pinned cross-tie DW to fit between the triangles could give rise to the observed vertical $\sigma'$ variations.

The comparison between the two triangle sizes further shows that halving $b$ from $5$ to $2.5~\mu$m nearly doubles the accessible coercive-field range. This scaling behavior highlights the enhanced tunability of domain-wall depinning at smaller length scales and suggests a promising route toward more precise control of domain-wall propagation.

More generally, a common trend emerges across all the cooperative arrays, including E-type arrays with variable $b$,  when one considers not only the available length for the domain wall at the bottom of the triangles (i.e., $bx$) but also the available length at the top of the triangles, which coincides with the horizontal periodicity of the arrays, $p=bx + b$. Lower panel of Fig.~\ref{fig:Pinningfield} illustrates the dependence $H_c= C/bx + E p $, with $H_c bx$ plotted against the product of the periodicity $p$ and the horizontal triangle separation $bx$   (i.e. $pbx$), with both axes displayed on a logarithmic scale. Despite some scatter of stochastic origin, the data follow an overall linear trend, as highlighted by the linear fit included in the log-log figure in dark red ($H_cbx= 3.89 + 0.01 p bx $, $R^2 = 0.75$).  Although the competing arrays provide fewer data points,  the model still yields a plausible fit that is consistent with the overall trend (see green symbols and line in Fig.~\ref{fig:Pinningfield}). This linear dependence of $H_c bx$ on $p bx$ can be phenomenologically described by a non-zero intercept $D=Ep$, suggesting a possible connection between variations in DW surface energy and the effect of array periodicity on the internal DW structure (i.e., $D\propto p\propto\sigma'$).

\section{Conclusions}

We have shown that domain-wall propagation in Co$_{0.86}$Zr$_{0.14}$ amorphous films patterned with triangular-hole arrays is governed by the interplay between hole orientation and density gradients. When the gradient and triangular-hole asymmetry cooperate, the preferred propagation direction is reinforced, producing a stronger ratchet effect. When they compete, the gradient can overcome the local geometric asymmetry, producing an inverse ratchet and opening experimental access to backward domain-wall propagation. The reduced dispersion of depinning fields in density-graded arrays further indicates that spatially varying pinning landscapes can improve the reproducibility of domain-wall motion.

Our results show that the depinning field in density-graded arrays is governed primarily by the local domain-wall length, as predicted by both the elastic and rigid DW classical descriptions, with the latter providing a better account of the observed forward/backward asymmetry. This agreement is consistent with the relatively rigid cross-tie structure of the DWs observed in XMCD-PEEM, which also suppresses cross-ratchet effects at large hole separations. At the same time, the depinning field is also influenced by the array periodicity, which affects the internal V–AV structure of the cross-tie DW. An empirical scaling law incorporating both geometrical parameters successfully captures the main experimental trends and provides a practical guideline for array design.

More broadly, these results demonstrate that domain-wall dynamics can be tuned through the combined engineering of the pinning landscape and the internal structure of the wall, beyond local geometric asymmetry alone. The resulting scaling relations provide a framework for designing magnetic arrays in which domain walls can propagate in either direction and be pinned at selected locations, with potential relevance to energy-efficient spintronic, memory, sensing, and neuromorphic devices.

\section{Methods}

\label{par:empezar}

The investigated arrays were defined on approximately 40~nm-thick amorphous Co$_{0.86}$Zr$_{0.14}$ thin films deposited on Si(100) substrates by co-sputtering Co and Zr targets. During deposition, an oblique incidence angle was used to induce a uniaxial magnetic anisotropy with an in-plane easy axis (EA). Magnetic characterization of the thin films, including angular-dependent hysteresis loops, was performed using a Quantum Design MPMS3 SQUID magnetometer and a NanoMOKE3 (Magneto-Optic Kerr Effect, Durham Magneto Optics Ltd). 

The films were coated with PMMA resist and then patterned by electron-beam lithography (Raith150two) using an acceleration voltage of 10~kV, a beam current of 0.24~nA, and a dwell time of 0.4~$\mu$s. The magnetic pattern was subsequently defined using an ion milling system (4Wave) with an argon flow rate of 15 standard cubic centimeters per minute, an acceleration voltage of 300~V, a beam current of 100~mA, and a beam voltage of 800~V, for a duration of 5~min and 30~s. Subsequent X-ray diffraction measurements confirmed the absence of recrystallization in the films.

Kerr microscopy images of complete arrays were acquired using a wide-field EvicoMagnetics microscope in longitudinal configuration. An in-plane magnetic field was applied parallel to the easy axis. Domain wall propagation was recorded live during field sweeps. The position of the domain walls within the arrays was monitored as a function of the applied magnetic field, from which local depinning (or coercive) fields and the magnetization of the complete arrays were obtained. Only one hysteresis loop per array is included in the paper to ensure consistent statistics across all arrays; additional loops measured for some arrays yielded equivalent results.

High-resolution domain configurations and domain wall (DW) morphologies were examined using X-ray PhotoEmission Electron Microscopy (XPEEM) at the BL24 – CIRCE beamline of the ALBA synchrotron. X-ray Magnetic Circular Dichroism (XMCD) contrast was obtained at the Co $L_3$ absorption edge, providing sensitivity to the magnetization projection along the X-ray beam direction. Measurements were conducted using a magnetic sample holder capable of applying in-plane magnetic fields during imaging  \cite{Foerster2016}.
magnetization maps have been  reconstructed by combining XMCD-PEEM images acquired at different azimuthal angles. The relative geometric distortions between images were corrected using a state of the art  deep-learning dense feature matching model based on multiple neural-network architectures  \cite{edstedt2025romav2, Zubizarreta2026}, allowing all XMCD intensity maps to be projected onto a common geometry before combining them into the final vector magnetization image.

\section{Supplementary material}

See the supplementary material for details on the unpatterned thin film characterization.

\section*{Acknowledgements}

This work was supported by MICIU/AEI/10.13039/501100011033 under grants PID2020-115159GB-I00, PID2022-136784NB, CEX2023-001286-S, PID2024-159457OB-C21, PID2024-157172-NB-I00, PID2021-122980OB-C54, PID2024-157112OB-C55, and PID2025-167788NB; the Government of Aragon (E12-23R RASMIA); and Agencia SEKUENS (Asturias) (IDE/2024/000678), with support from the European Union through FEDER funds. X. Z-I.: staff hired under the Generation D initiative, promoted by Red.es, an organisation attached to the Ministry for Digital Transformation and the Civil Service, for the attraction and retention of talent through grants and training contracts, financed by the Recovery, Transformation and Resilience Plan through the European Union's Next Generation funds. X-PEEM experiments were performed at BL24 - CIRCE beamline at ALBA Synchrotron with the collaboration of ALBA staff. Authors would like to acknowledge the use of Servicio General de Apoyo a la Investigación-SAI, Universidad de Zaragoza. 
We thank Rubén Martínez Cantín for inspiring the image registration approach used to generate the 2D magnetization map.

\bibliography{AAAreferences}

\newpage

\section*{Supplementary material}
\subsection*{Thin film characterization}

Magnetization measurements, presented in Figs.~\ref{fig:SQUID}~and~\ref{fig:MOKE}, show that the thin film has a saturation magnetization of $M_S = 7 \cdot 10^{5}$~A/m and a coercive field of $H_c = 5 \cdot 10^{2}$~A/m and that the film exhibits the expected behavior for a well-defined uniaxial anisotropy: the reduced remanence follows \(M_R/M_S \propto \cos(\theta)\), while the coercivity displays the anticipated angular dependence, \(H_c \propto 1/\cos(\theta)\), over a wide range of angles \cite{Chikazumi1978, Perez-Junquera2004, perez-junquera2006}.

\setcounter{figure}{0}
\renewcommand{\thefigure}{\Alph{figure}}

\begin{figure}[!htb]
\centering
\includegraphics[width=6cm]{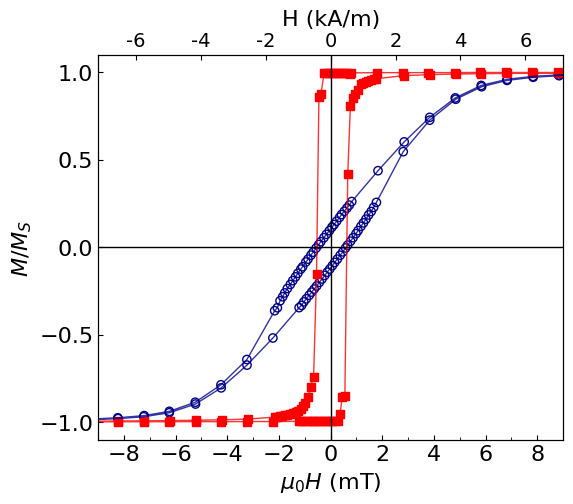}
\caption{SQUID hysteresis loops at the easy axis (filled squares) and the hard axis (hollow circles) of the unpatterned 40-nm-thick Co$_{0.86}$Zr$_{0.14}$ film.}
\label{fig:SQUID}
\end{figure} 

\begin{figure}[!htb]
\centering
\includegraphics[width=8.5cm]{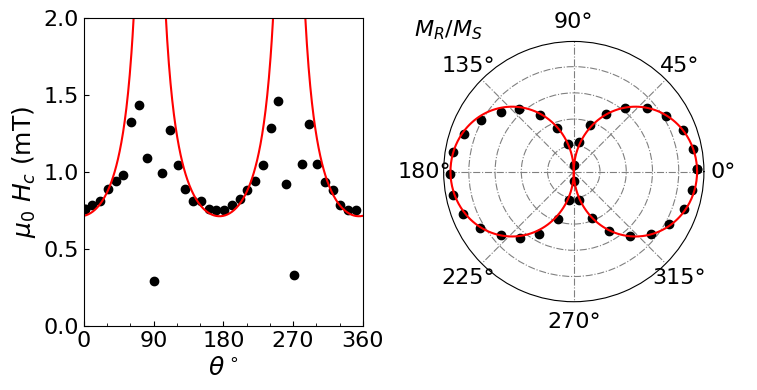}
\caption{MOKE measurements on the unpatterned film. Left: angular dependence of the coercive field (dots) with a $1/cos(\theta)$ dependence (line). Right: polar plot of the reduced remanence (dots) with a $\cos(\theta)$ fit (line).}
\label{fig:MOKE}
\end{figure} 

\begin{figure}[!p]
 \centering
 \includegraphics[width=8.5cm]{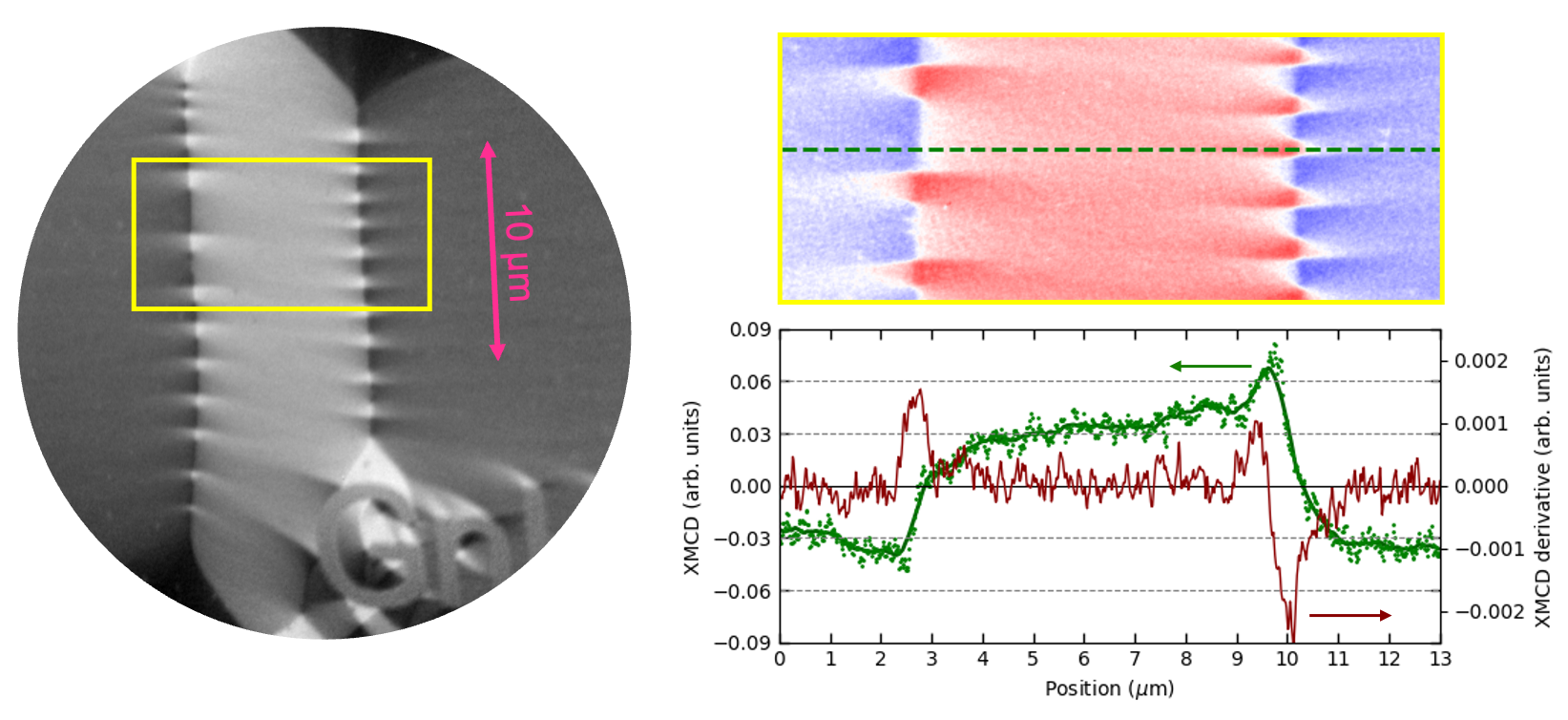}
 \caption{Left: XMCD image at the Co $L_3$ edge of the continuous film, shown in grayscale, revealing two parallel cross-tie domain walls. Top right: magnified view of the region enclosed by the yellow line in the left image, shown using a red--white--blue scale for positive, zero, and negative XMCD signal, respectively. Bottom right: XMCD signal (green dots) and its smoothed value (green line) along the green line marked in the top-right image. The line crosses the DW at a section where the wall exhibits a Néel-like configuration (left) and across an antivortex (right). The derivative of the green line is shown in dark red. The DW tail is approximately 2~$\mu$m at the first crossing and approximately twice this value at the antivortex.
 }
 \label{fig:DWwidth}
\end{figure}

Fig.~\ref{fig:DWwidth} shows an XMCD-PEEM image acquired at the Co $L_3$ absorption edge, displaying three magnetic domains in the continuous film separated by two parallel, oppositely oriented cross-tie-type DWs. The DW width is approximately 2~$\mu$m in the sections with a Néel-like configuration and nearly twice as large in the vicinity of a vortex or antivortex.

\clearpage

\end{document}